\documentclass[aps,prl,reprint,showpacs,superscriptaddress,amsmath,amssymb,floatfix]{revtex4-1}
\usepackage{graphicx}
\usepackage{graphics}
\usepackage{color}

\begin{document}



\title{Shear accelerated crystallization in a supercooled atomic liquid}

\author{Zhen Shao}
\affiliation{Department of Chemical and Environmental Engineering, Yale University, New Haven CT 06511}
\affiliation{Center for Research on Interface Structures and Phenomena, Yale University, New Haven CT 06511}

\author{Jonathan P. Singer}
\affiliation{Department of Chemical and Environmental Engineering, Yale University, New Haven CT 06511}
\affiliation{Center for Research on Interface Structures and Phenomena, Yale University, New Haven CT 06511}

\author{Yanhui Liu}
\affiliation{Center for Research on Interface Structures and Phenomena, Yale University, New Haven CT 06511}
\affiliation{Department of Mechanical Engineering and Materials Science, Yale University, New Haven CT 06511}

\author{Ze Liu}
\affiliation{Center for Research on Interface Structures and Phenomena, Yale University, New Haven CT 06511}
\affiliation{Department of Mechanical Engineering and Materials Science, Yale University, New Haven CT 06511}

\author{Huiping Li}
\affiliation{Center for Research on Interface Structures and Phenomena, Yale University, New Haven CT 06511}
\affiliation{Department of Mechanical Engineering and Materials Science, Yale University, New Haven CT 06511}

\author{Manesh Gopinadhan}
\affiliation{Department of Chemical and Environmental Engineering, Yale University, New Haven CT 06511}
\affiliation{Center for Research on Interface Structures and Phenomena, Yale University, New Haven CT 06511}

\author{Corey S. O'Hern}
\affiliation{Center for Research on Interface Structures and Phenomena, Yale University, New Haven CT 06511}
\affiliation{Department of Mechanical Engineering and Materials Science, Yale University, New Haven CT 06511}

\author{Jan Schroers}
\email[]{jan.schroers@yale.edu}
\affiliation{Center for Research on Interface Structures and Phenomena, Yale University, New Haven CT 06511}
\affiliation{Department of Mechanical Engineering and Materials Science, Yale University, New Haven CT 06511}

\author{Chinedum O. Osuji}
\email[]{chinedum.osuji@yale.edu}
\affiliation{Department of Chemical and Environmental Engineering, Yale University, New Haven CT 06511}
\affiliation{Center for Research on Interface Structures and Phenomena, Yale University, New Haven CT 06511}


\date{\today}

\begin{abstract}
 A bulk metallic glass forming alloy is subjected to shear flow in its supercooled state by compression of a short rod to produce a flat disc. The resulting material exhibits enhanced crystallization kinetics during isothermal annealing as reflected in the decrease of the crystallization time relative to the non-deformed case. The transition from quiescent to shear-accelerated crystallization is linked to strain accumulated during shear flow above a critical shear rate $\dot\gamma_c\approx 0.3$ s$^{-1}$ which corresponds to P\'{e}clet number, $Pe\sim\mathcal{O}(1)$. The observation of shear-accelerated crystallization in an atomic system at modest shear rates is uncommon. It is made possible here by the substantial viscosity of the supercooled liquid which increases strongly with temperature in the approach to the glass transition. We may therefore anticipate the encounter of non-trivial shear-related effects during thermoplastic deformation of similar systems.
\end{abstract}

\pacs{81.05.Kf,61.43.Fs,61.25.Mv,64.70.pe} 

\maketitle


The ability of flow fields to modify the structure and phase behavior of condensed matter has been well described in a variety of systems. Plastic strain in solid metals results in texture development \cite{Segal1995materials} while steady or reciprocating shear can orient microstructures in polymer melts \cite{Lagasse1976,Li_deJeu2003shear}, colloidal suspensions \cite{Chen1992,wu2009melting}, block copolymers \cite{Winey1993,Chen_Kornfield1997} and lyotropic surfactant mesophases \cite{Penfold1997shear}. Likewise, shear can suppress or enhance phase stability \cite{onuki1997phase} and in particular, shear enhanced crystallization has been experimentally observed in a broad range of materials.

The acceleration of crystallization in macromolecular systems under shear is understood to originate from the flow alignment of chains which reduces the entropy of the melt and biases the system towards crystalization. This effect is particularly acute during nucleation, and the large relaxation time of entangled polymer melts allows for this behavior at relatively low shear rates \cite{Elmoumni2006}. Similarly, shear-induced crystallization in colloidal systems occurs in regimes of flow where the suspension microstructure can be significantly perturbed by the flow field. The transition to such a regime is described by the dimensionless P\'{e}clet number, $Pe =\dot\gamma \tau_d$ which captures the relative importance of advective and diffusive mass transport, with $\dot\gamma$ the shear rate and $\tau_d$ the timescale for particle diffusion. Ordered packed particle layers in hard sphere suspensions are often observed at modest shear rates where $Pe >1$ \cite{Vermant2005flow,wu2009melting,Cheng_Cohen2011}. Such structures may accelerate nucleation, while the shear field can also enhance the growth rate of existing nuclei. Conversely, excessive shear flow can ``shear melt'' colloidal crystals.

As canonical examples, mesoscopic systems such as polymer melts and colloidal suspensions highlight the balance of timescales that defines the emergence of shear-influenced crystallization, and allow quantitative assessment of this effect. By comparison, apart from recent molecular dynamics studies \cite{Mokshin2009,Mokshin2010,Kerrache2011,Mokshin2011,Mokshin2013}, this topic has remained largely unexplored in atomic systems. Experimental progress has been impeded due to the practical difficulties associated with the prohibitively high shear rates needed to achieve $Pe\sim\mathcal{O}(1)$ for fast-relaxing atomic liquids and melts. To date, shear-induced deviations of the crystallization of amorphous metals from the quiescent case have not been observed.

Here, we present systematic, quantitative evidence that shear accelerates crystallization of an atomic melt at substantially lower shear rates. We consider a bulk metallic glass (BMG) forming alloy and identify a critical shear rate of $\dot\gamma$=0.3 $\mathrm{s^{-1}}$ above which substantial shear-related effects can be observed in the kinetics of isothermal crystallization subsequent to flow, and below which the material displays behavior similar to that of the quiescent case. Using the Volgel-Fulcher-Tammann (VFT) form for the temperature dependence of the viscosity we correlate this critical shear rate with $Pe\sim\mathcal{O}(1)$. The ability to observe shear-induced effects at experimentally accessible shear rates is linked to the highly viscous nature of the melt in the supercooled state. The modest shear rates at which flow influences crystallization suggests that shear-accelerated crystallization must be properly accounted for in BMG forming operations.


\begin{figure}[t]
\begin{minipage}[t]{16mm}
\vspace{0pt}
\includegraphics[width=16.5mm, scale=1]{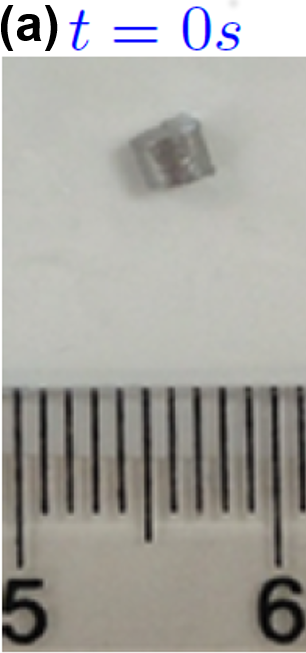}\\
\end{minipage}%
\begin{minipage}[t]{16mm}
\vspace{0pt}
\includegraphics[width=16.5mm, scale=1]{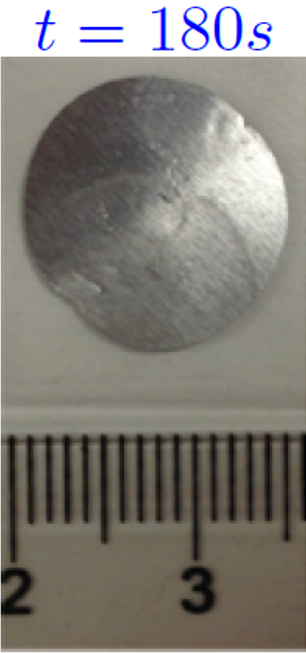}
\end{minipage}%
\begin{minipage}[t]{52mm}
\vspace{0pt}
\includegraphics[width=48mm, scale=1]{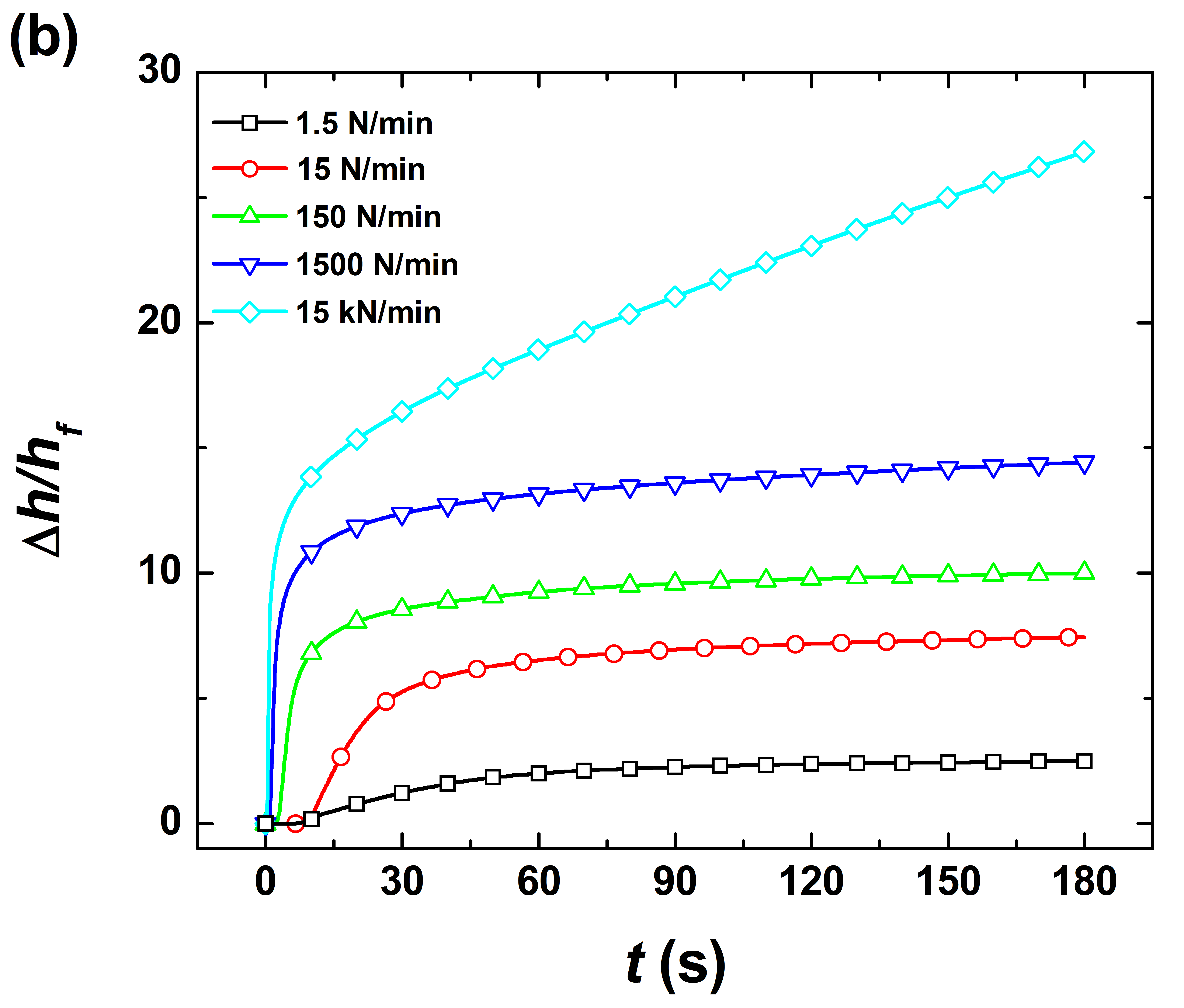}
\end{minipage}%
\caption{(Color online) (a) A short BMG rod is pressed to a flat disc using a loading rate of 15kN/min for 180s at 270 $^{\circ}$C. (b) Normalized compression as a function of time for selected loading rates. $h_f$ is the final thickness of the disc.}
\label{experimental_schematic}
\end{figure}

The system under investigation here is a Pt-based alloy, $\mathrm{Pt_{57.5}Cu_{14.7}Ni_{5.3}P_{22.5}}$. Samples were subjected to shear \textit{via} deformation during uniaxial compression from a rod-like pellet to a thin disc. In a forming experiment, a BMG rod is placed between two flat platens that are maintained at a constant temperature of 270 $^{\circ}$C as depicted in Fig. \ref{experimental_schematic}. For this alloy, this temperature is in the supercooled liquid regime, which is bounded by the glass transition temperature, $\mathrm{T_g}$ = 235 $^{\circ}$C, and the crystallization temperature, $\mathrm{T_x}$ = 305 $^{\circ}$C. In the supercooled state, the alloy is a sluggish liquid with a viscosity $\eta$ ranging from $\mathrm{10^6-10^{12}}$ Pa.s \cite{Legg_Schroers2007}. A 3 minute load profile with constant loading rate ranging from 1.5 to 15000 N/min was applied. This resulted in a reduction in thickness ranging from a factor of $\approx$ 3X-30X depending on the loading rate. In all cases the shear strain rates exceeded the compressional strain rates for the majority of the deformation, $\dot\gamma=\dot\varepsilon_{xz}>4\dot\varepsilon_{zz}$ especially at the advancing interface of the BMG where the shear and, thereby, shear-induced effects were the largest magnitude.

\begin{figure}[t]
\includegraphics[width=75mm, scale=1]{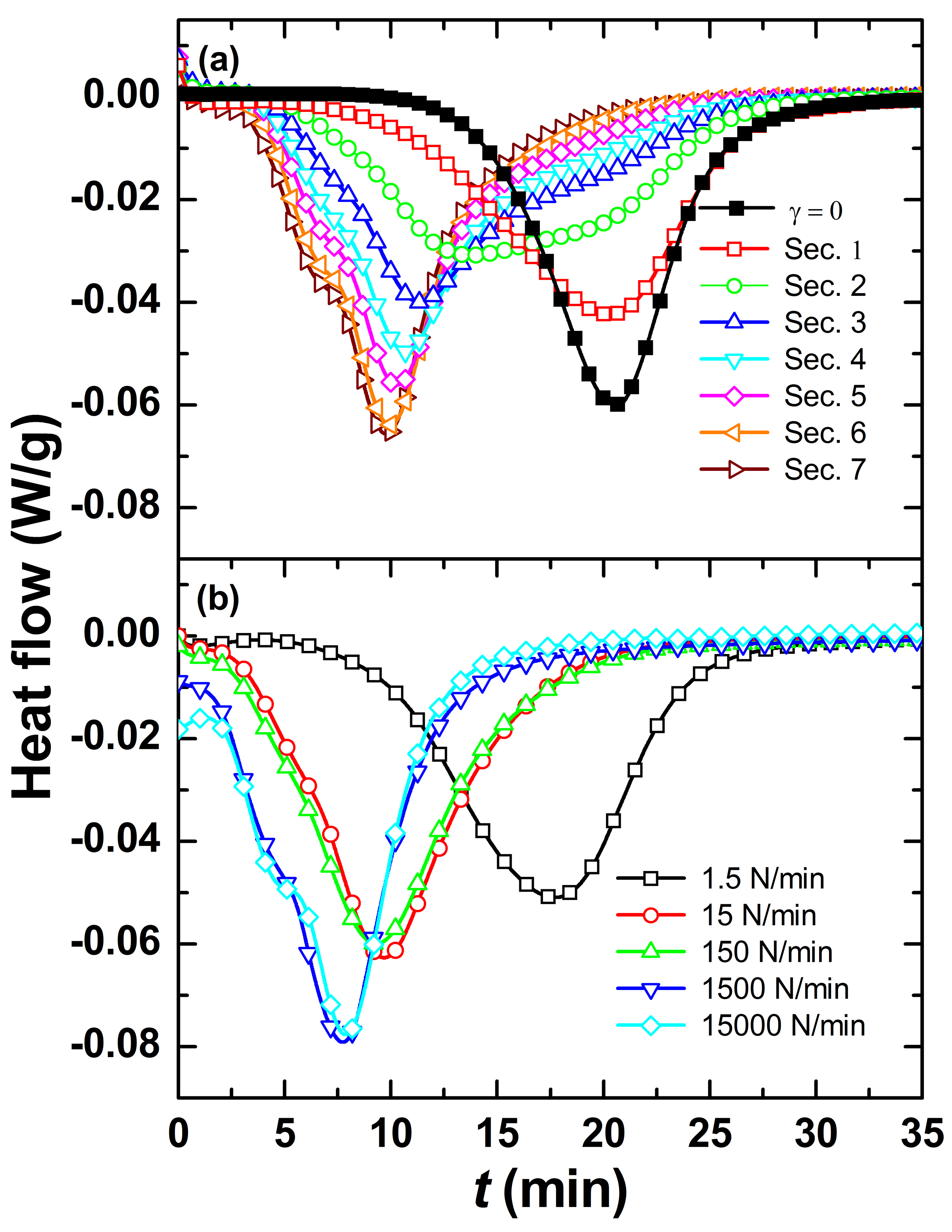}
\caption{(Color online) Isothermal thermographs of pressed discs. (a): Circular annuli from Sample 6 shownig radial dependence (1: center; 7: edge). Reference quiescent sample ($\gamma=0$) is also shown. (b): Wedge-shaped samples from discs produced using different loading rates as indicated. Note that samples in (a) were slightly thicker than those in (b) resulting in lower maximum shear rates and therefore larger values of the minimum (radially dependent) $t_x$.}
\label{dsc_data}
\end{figure}

The crystallization kinetics were characterized by differential scanning calorimetry (DSC), with heat flow recorded during isothermal annealing at 270 $^{\circ}$C. Data were collected for different radial positions for a single loading profile (15 kN/min, Fig. \ref{dsc_data}a) and for wedge-shaped disc sections prepared at the different loading rates in Table 1. In the first case, pressed discs were sectioned into several circular annuli which were evaluated separately. In the second case, wedge-sections were individually characterized to provide a description of crystallization behavior integrated over the radially-dependent shear rates produced by each loading profile. For experimental convenience and more accurate detection, the crystallization time $t_x$ is defined as the time of maximum heat flow. $t_x$ defined in this manner corresponds to the time when the extent of crystallization is $\sim$50\% due to the near-symmetric shape of the crystallization peak in the thermogram.

\begin{table}
\begin{tabular}{|c|c|c|c|c|}
\hline
\hline
Sample & Loading rate (N/min) & $h_f$ ($\mu$m) & $t_x$(min) \\
\hline
\hline
1 &  0  & 2500 & 20 \\
2 &  1.5  & 724 & 17.7 \\
3&15&219 & 9.7\\
4&150&165 & 9.3\\
5&1500&106 & 7.7\\
6&15000&75 & 7.9\\
\hline
\hline
\end{tabular}
\caption{Loading rates, final thickness $h_f$, and crystallization times $t_x$. Corresponding DSC data are in Fig. \ref{dsc_data}b.}
\label{exp_table}
\end{table}

The quiescent sample is exposed to the same thermal treatment (180s at 270 $^{\circ}$C) but not subjected to any deformation and therefore provides a baseline for the crystallization behavior. Data are shown in Fig. \ref{dsc_data}a for $\dot\gamma=0$. Crystallization for the quiescent case occurs with a highly reproducible $t_x$ of 20 minutes, consistent with published data \cite{Legg_Schroers2007} after properly accounting for the 180s latent period.\cite{Schroers2000} For the pressed disc, the $t_x$ for material sampled from the center of the disc out to $R/7$, (where $R$ is the radius of the disc) is also ~20 minutes (purple trace in Fig. \ref{dsc_data}a), effectively indistinguishable from the quiescent sample. Data taken from successive circular annuli show a progressive decrease of $t_x$ to a minimum of ~10 minutes. The effects of different shear rates produced by varying the loading rate are shown in Fig. \ref{dsc_data}b. $t_x$=~18 minutes for samples pressed at the lowest loading rate (1.5N/min) while material deformed with the highest rate (15 kN/min) crystallizes after only ~7 minutes of additional heating, about three times faster than the quiescent case.

As discussed above, during deformation the sample experiences different shear rates as a function of radial position and the applied loading rate. The data unambiguously demonstrate that $t_x$ is a strong function of position and loading rate and thus we can conclude that the crystallization kinetics are accelerated by shear. One may speculate that heat evolved during deformation of the system could contribute; however the rate of heat transport out of the material significantly exceeds the rate of generation, excluding this effect (see Supplemental Material \cite{zhaoPRESupp}). We propose instead that crystallization is accelerated due to local ordering in volume elements of the material where the shear rate exceeds a critical shear rate $\dot\gamma_c$ where the structure of the liquid is dictated by advection.

Under quiescent conditions, mass transport occurs by diffusion alone, with the characteristic timescale $\tau_d$ set by the diffusivity $D$ and the characteristic atomic length scale $a$, $\tau_d=a^2/D$. The advective timescale due to shear is given by the inverse shear rate, $\tau_d=\dot\gamma^{-1}$ and so  $Pe=\dot\gamma a^2/D$. We define a critical shear rate $\dot\gamma_c$ as that at which the atomic transport by advection dominates diffusive transport. For $Pe\sim\mathcal{O}(1)$ and $\dot\gamma>\dot\gamma_c$, the microstructure of the melt is therefore dictated by the advection and the crystallization kinetics become a function of the shear rate. The critical shear rate can be approximated in terms of the melt viscosity through the Stokes-Einstein relation $D=k_BT/3\pi\eta a$. The melt viscosity is assumed to exhibit a VFT temperature dependence, $\eta=\eta_0\exp[F^*T_0/(T-T_0)]$ where $F^*$, $\eta_0$ and $T_0$ are fitted empirical constants. The temperature dependence of the critical shear rate for flow-dominated crystallization is then given by Eq. \ref{eq:crit_shear_rate}.

\begin{equation}
\dot\gamma_c=\frac{k_BT}{3\pi\eta_0 a^3}\exp\left[-\frac{F^*T_0}{T-T_0}\right]
\label{eq:crit_shear_rate}
\end{equation}

\begin{figure}[t]

\includegraphics[width=60mm, scale=1]{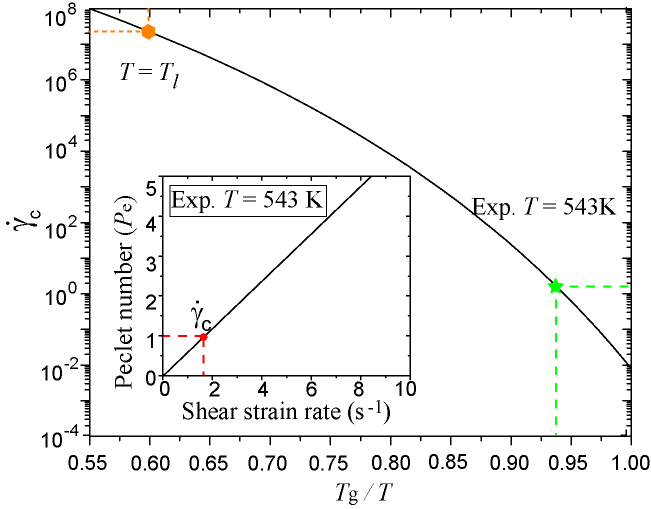}
\caption{(Color online) Temperature dependent critical shear rate, Eq. \ref{eq:crit_shear_rate}.}
\label{temp_crit_rate}
\end{figure}

For $\mathrm{Pt_{57.5}Cu_{14.7}Ni_{5.3}P_{22.5}}$, $\eta_0$=4$\times$10$\mathrm{^{-5}}$Pa.s, $T_0$=336 K and $F^*$ = 16.4 \cite{Legg_Schroers2007,Gallino2010}. Based on atomic radii of 177, 145, 149 and 98 pm for Pt, Cu, Ni and P, we calculate a volume weighted average atomic size (diameter) of $a$ = 0.32 nm, and calculate $\dot\gamma_c(T)$, Fig. \ref{temp_crit_rate}. For the experimental temperature of 543 K $\dot\gamma_c \approx 1.7$.

Further consideration of the above treatment requires establishing the relationship between the applied loading rate and radially-dependent local shear rates, as the strain accumulated above $\dot\gamma_c$ should provide a strong correlation with the kinetics. The spatial dependence of the instantaneous shear rate in a disc subjected to compression by no-slip parallel plates can be written analytically as a function of the compression rate and the original dimensions (Supplemental Material \cite{zhaoPRESupp}). A proper description however must account for the radially outward physical transport of volume elements during flow as the disc is flattened. We therefore use a finite element simulation (Supplemental Material \cite{zhaoPRESupp}) to fully determine the mechanical history of the samples. This allows the radial dependence of accumulated strain during flow to be quantitatively determined.

\begin{figure}[t]
\hspace{0mm}
\includegraphics[width=70mm, scale=1]{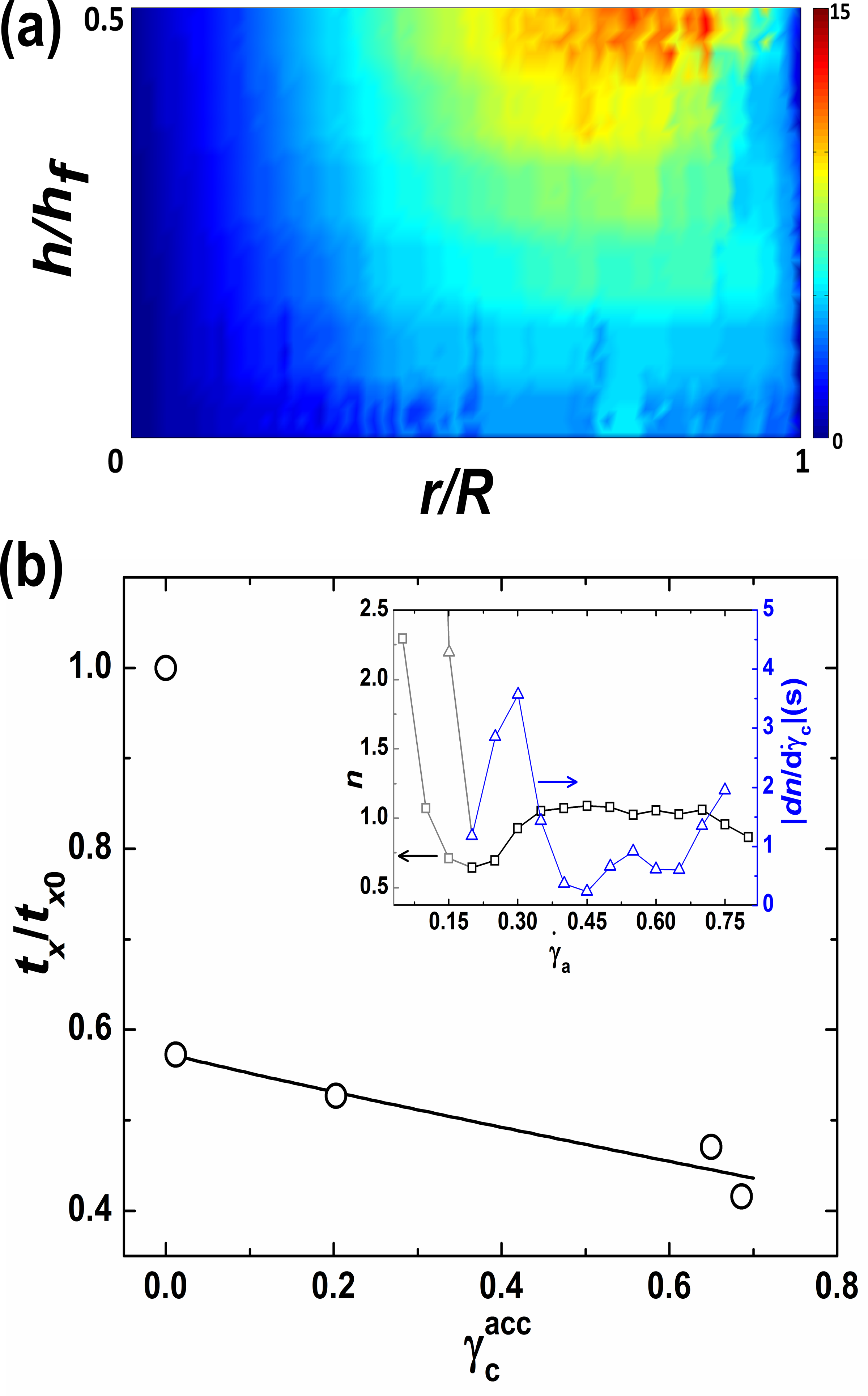}
\caption{(Color online) (a) Accumulated strain from finite element calculation. $h=0$ corresponds to the mid-plane of the disc. (b) Dependence of normalized crystallization time $t_x/t_x0$ on simulated volume-averaged accumulated critical strain $\gamma_c^{ac}$ for $\dot\gamma_c$=0.3 $\mathrm{s^{-1}}$ corresponding to Sample 4. $tx_0$ is the crystallization time of the pristine material. Inset: radial dependence of $\gamma_c^{ac}$. Location ($r$) is normalized by the disc radius, $R$. Lines are drawn as guides.}
\label{accumulated_strain}
\end{figure}

We calculate the volume-averaged accumulated strain that fluid elements at a given radial position encountered throughout their flow history. Specifically we consider strain accumulated only when the fluid parcels experience strain rates larger than $\dot\gamma_c$, Eq.\ref{eq:accumulated_strain}. Data from finite element calculations employing simulated loading rates of Sample 4 are used to determine $\dot\gamma_c$ by performing a sensitivity analysis. The distribution of accumulated strain is shown in Fig. \ref{accumulated_strain}a. Although the instantaneous shear rates are largest at the periphery of the disc, the accumulated strain displays a different distribution as the volume elements at the peripheries are constantly refreshed due to the generation of new surface during the deformation of the disc. The maximum accumulated strain therefore occurs slightly inward of the maximum in the instantaneous shear rate.

We consider the dependence of $t_x$ on the strain $\gamma_a^{ac}$ accumulated above a given rate $\dot\gamma_{a}$, with accumulated strain averaged over each of 4 annuli that collectively represent the full radial extent of the sample consistent with the experimental results. The gradient $n(\dot{\gamma_a}=\mathrm{d}t_x/\mathrm{d}\dot\gamma_a^{ac}$ exhibits a sigmoidal shape. $\dot\gamma_c$ is defined as the point at which there is a maximum in the rate of change of this dependence as a function of $\dot\gamma_a$, signaling the point at which the sensitivity of the dependence is greatest, that is $\dot\gamma_a$ for which $\mathrm{d}n/\mathrm{d}\dot\gamma_a$ exhibits a first (non-trivial) maximum. In this manner we estimate $\dot\gamma_c$=0.3 s$^{-1}$, inset Fig. \ref{accumulated_strain}b ($\gamma^{ac}_{c}(r,z)$ in Supplemental Material \cite{zhaoPRESupp}). $t_x$ shows a marked dependence on $\gamma^{ac}_{c}$, with a linear decrease over a broad range of accumulated critical strain after a sharp decrease from the quiescent value, Fig. \ref{accumulated_strain}b.

\begin{equation}
\gamma_c^{ac}=\int\displaylimits_{t:\dot\gamma>\dot\gamma_c}\int\displaylimits_0^{h'} \int\displaylimits_0^R \dot\gamma(r,h,t)\,r\,\mathrm{d}r\,\mathrm{d}h\,\mathrm{d}t
 \label{eq:accumulated_strain}
\end{equation}

Our experimentally determined $\dot\gamma_c$=0.3 s$\mathrm{^{-1}}$ differs from the predicted value of 1.7 for $Pe=1$ based on Eq. \ref{eq:crit_shear_rate}. The lack of exact agreement is not surprising however given the uncertainty associated with the parameters in the VFT description of the dynamics and the exponential dependence on two of these parameters, $F^*$ and $T_0$. Additionally, estimating the diffusivity using the Stokes-Einstein equation may represent an oversimplification \cite{Hodgdon1993,Geyer1996}. The correlation between increasing accumulated critical strain and rate of crystallization is intuitive in the context of the earlier discussion. Similar observations have been made for polymer melts crystallized during or subsequent to shear, with similarly strong correlations between the normalized crystallization rate and the accumulated strain or applied shear rate \cite{Acierno_Winter2003,Elmoumni2006}. However, simple models which can correctly account for the observed behavior are not available. The persistence of this situation despite the practical significance of flow-induced crystallization in melt-processed commodity-scale polymers such as polyolefins reflects the complexity of the underlying phenomena.

Consideration of crystallization and aggregation in colloidal systems may provide a useful framework for interpreting our results however, by examining the role of shear in effectively decreasing the activation barrier for nucleation \cite{Blaak2004} or aggregation \cite{Zaccone2009}. In the latter case, Zaccone \textit{et al.} provide an expression for modification of aggregation kinetics $~\exp(-V/kT + \alpha Pe)$ that may reasonably be extended to atomic systems using the appropriate potentials, where $V$ represents an activation barrier for nucleation and $\alpha$ depends on the flow geometry. In this framework, the transition to a shear-dominated regime occurs for $Pe\geq(1/\alpha)(V/kT)$. The exponential dependence suggests a sharp transition to shear dominated kinetics though in the present case any such sharpness is subject to smearing by the non-uniform shear history of fluid elements as they are convected during deformation. Experiments using a constant shear-rate geometry would enable these arguments to be more rigorously evaluated for the supercooled melts studied here.

\begin{figure}[ht]
\includegraphics[width=85mm, scale=1]{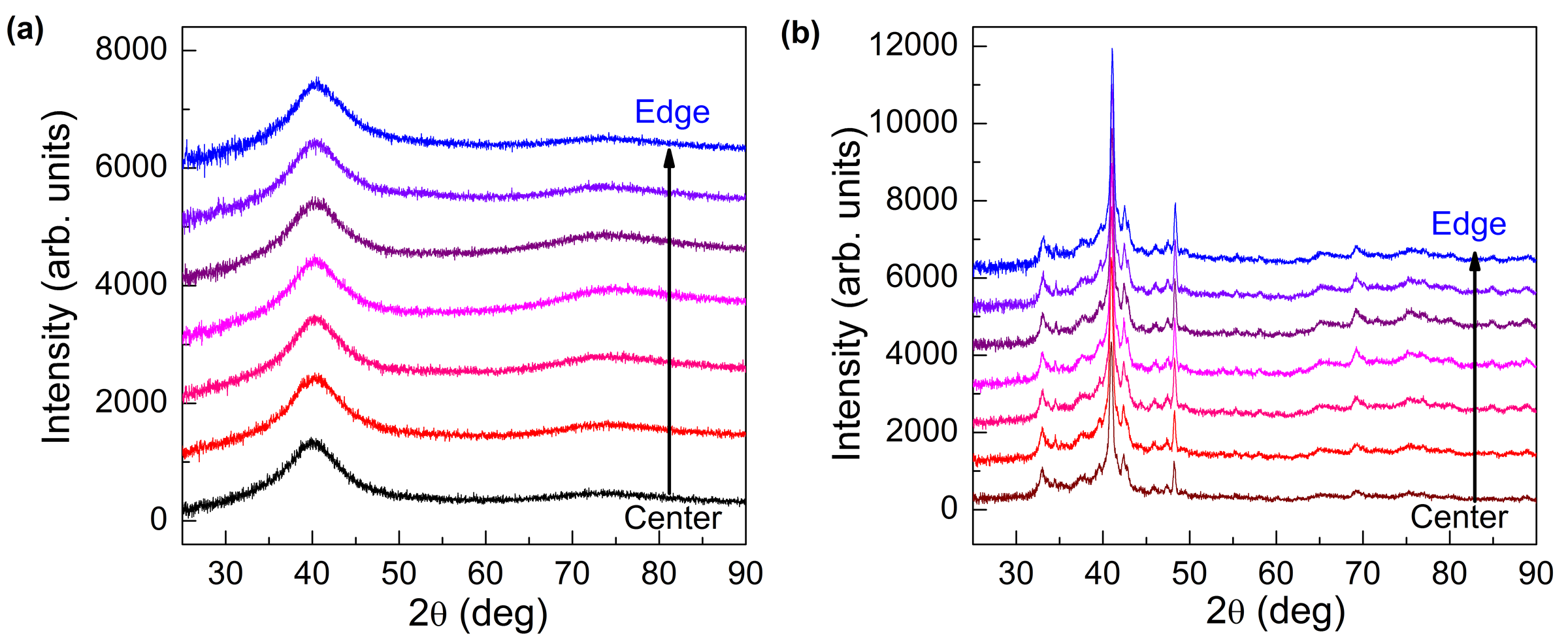}
\caption{(Color online) XRD of disc pressed at 15 kN/min loading rate. Data offset vertically for clarity. (a) Sample immediately after pressing. There is only one broad peak indicating the material is amorphous. (b) Pressed disc after isothermal crystallization. There are no discernable differences in grain size (peak widths) suggesting that no crystallization occurs during deformation of the sample.}
\label{xrd}
\end{figure}

In the present case we have shown that an atomic melt may also display strong shear-induced crystallization behavior, which we quantitatively link to flow above a critical shear rate corresponding to $Pe\sim\mathcal{O}(1)$. We propose that shear-driven local ordering is responsible for this display, in much the same manner as observed in mesoscopic systems such as colloidal suspensions \cite{wu2009melting}. X-ray diffraction (XRD) of as-pressed discs reveals a broad amorphous hump centered at $\mathrm{2\theta=40^{\circ}}$ that is characteristic of the glass, with no discernable differences in the structure of the pressed discs as a function of radial position (Fig. \ref{xrd}). XRD of fully crystallized material likewise does not display any significant differences radially of peak width, and therefore crystallite size as interpreted by Scherrer analysis. The difficulty of detecting nanoscale crystallites in a bulk amorphous matrix prevents us from concluding that there are no such crystallites present. However from these XRD experiments as well as cursory TEM investigations, we have no indication of crystallite formation during shear. This is consistent with the enhancement of nucleation and growth during subsequent isothermal annealing as being the result of subtle changes in the structure of the supercooled liquid. We speculate that the multi-component nature of the amorphous alloys may also play a role in this regard. The complex composition that is often a prerequisite for the suppression of crystallization in bulk glass formers \cite{greer1993confusion,Peker1993highly,kelton1998new} requires non-polymorphic crystallization in which the composition of the crystal is different than that of its surrounding melt. In addition to local ordering driven by advection, for non-polymorphic crystallization, shear flow may act to encourage local compositional heterogeneities due to polydispersity in the size of the constituent atoms. We note that accelerated crystallization kinetics were also observed in another BMG-former, $\mathrm{Zr_{44}Al_{10}Ti_{11}Cu_{10}Ni_{10}Be_{25}}$, as described in Supplemental Material \cite{zhaoPRESupp}.


In conclusion, we observed enhanced crystallization kinetics in a metallic glass subsequent to shear flow in its supercooled liquid state. We interpret this as a transition from diffusion to advection-dominated transport at a critical shear rate. A pairing of finite element calculations and experimental data permits estimation of the critical shear rate using a minimal set of assumptions, and yields rough agreement with simple estimates based on the P\'{e}clet number. The strain rates in our experiments are comparable with deformation rates encountered during thermoplastic forming \cite{Schroers2011thermoplastic}. We therefore anticipate that significant shear-crystallization effects will be relevant during BMG thermoplastic forming, particularly in the vicinity of $T_g$ where the strong divergence of viscosity drives $\dot\gamma_c$ down to small values. This may assume added significance in confinement where finite size effects come into play \cite{Gopinadhan2013finite,Shao2013size}. It is likely that crystallization in shear bands during deformation of metallic glasses \cite{Chen1994,Kim2002,lee2008nanocrystallization} occurs subject to the same considerations detailed here. Indeed, the comparison to shear induced crystallization in polymers by enhanced atomic diffusivity through shear has been advanced for this scenario \cite{Kim2002}.

\begin{acknowledgments}
This work was supported by NSF MRSEC DMR-1119826. Facilities use was supported by YINQE. J.S., Z.L., H.L and Y.L contributed equally to this work.
\end{acknowledgments}

\bibliographystyle{apsrev}
\bibliography{shear_enhanced_crystallization}


\section{Supplementary Material}

\section{Analytical Solution}

Equation \ref{eq:shear_field} provides an analytical expression for the instantaneous shear field in a disc subjected to compression by no-slip parallel plates as a function of the compression rate and the original dimensions.\cite{Engmann2005}  This quantifies the instantaneous flow field with maximum shear rates at the rims of the disc as shown in Figure \ref{flow_field}.

\begin{equation}
\dot\gamma(r,z)=\frac{3\dot h}{\sqrt{2}h}\left[(r/h)^2(z/h)^2+3\left(1-(z/h)^2\right)^2\right]^{1/2}
\label{eq:shear_field}
\end{equation}

\section{Finite Element Simulation}

Simulations were performed using the finite element method (FEM) in a similar manner as used in prior studies of squeeze flow \cite{Mavridis1992,Adams1997}. For this study, the commercial FEM software Marc was used, and the simulation was conducted for a Newtonian fluid cylinder with typical dimensions for a pressed piece ($h_i$=$d_i$=2.5 mm). Simulation was conducted with no-slip (appropriate for the creep flow of BMG) moving boundaries with a fixed velocity, which is one of the two ways that these problems are simulated (other being constant load) and terminated when the disk reached a height similar to that experimentally observed. Despite the use of adaptive remeshing, the massive deformations ($h_i/h_f\sim2-20$) involved eventually led to divergence in the simulation below final thicknesses of 125 $\mu$m, which led to the most accurate comparison being with the lowest of the sectioned loading rates (Sample 4, 150 N/min loading rate, $h_f$=165 $\mu$m). To track the shear rate experienced by a given element through the various remeshings and also correct for the fact that the actual experimental conditions were constant load rate rather than constant wall velocity, a MATLAB wrapper was written that at every time step (i) changed the wall velocity to match experimentally observed values (a valid scaling for a Newtonian fluid), (ii) tracked the accumulated shear and shear rate as a function of position, and (iii) mapped these onto the appropriate mesh point for the next deformation step.

\begin{figure*}[h]
\includegraphics[width=160mm, scale=1]{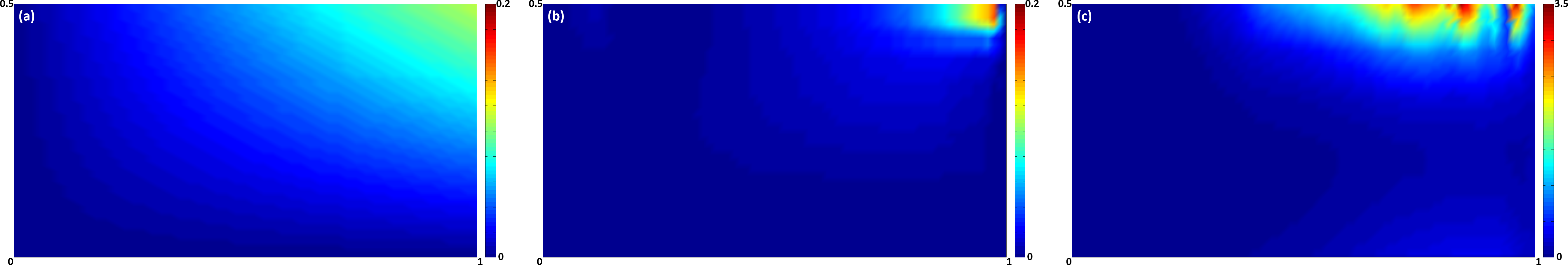}
\caption{(a) Analytical solution of shear rate distribution at t=3 min using Equation \ref{eq:shear_field} where $z_f$ and $r_f$ are the final height and radii of the disc, under the assumption of a fixed compression rate $\dot h$. $h=0$ corresponds to the mid-plane of the sample. (b) Shear rate distribution from finite element calculation. (c) Distribution of accumulated critical strain for $\dot\gamma_c$=0.3 s$\mathrm{^{-1}}$.}
\label{flow_field}
\end{figure*}

The spatial dependence of the instantaneous shear rate from the deviatoric component of the strain rate in a disc subjected to compression by parallel plates under no-slip boundary conditions can be written analytically as a function of the compression rate and the original dimensions, Equation \ref{eq:shear_field}.\cite{Engmann2005} Instantaneous shear fields produced using the analytical expression and the FEM, and accumulated critical strain are shown in Figure \ref{flow_field}.
\begin{equation}
\dot\gamma(r,z)=\frac{3\dot h}{\sqrt{2}h}\left[(r/h)^2(z/h)^2+3\left(1-(z/h)^2\right)^2\right]^{1/2}
\label{eq:shear_field}
\end{equation}

\section{Heat evolved during flow}

Considering the plastic work contribution to heat generation, the heat conduction equation is modified to
\begin{equation}
\dot{T} = \alpha {\nabla}^2T+\frac{\beta\sigma\cdot{\dot{\varepsilon}}^p}{\rho c_p}
\label{A1}
\end{equation}
where we neglected the thermoelastic effect for Newton fluid and $\frac{\beta\sigma\cdot{\dot{\varepsilon}}^p}{\rho c_p}$ is the generated heat due to irreversible mechanical work \cite{Taylor1934,Mason1994}. $\sigma$ is the stress, $\dot{\varepsilon}^p$ is the plastic strain rate, $\beta$ is the work rate to heat rate conversion fraction and it is simply assumed to be a constant in the range 0.85-1.00 \cite{Mason1994}. For squeezing flow between parallel plates with no-slip boundary condition, the generalized shear rate is given by Eq. \ref{A2} \cite{Engmann2005}
\begin{equation}
\dot{\gamma}(r,z)=\frac{3\dot{h}}{\sqrt{2}h}\sqrt{\left(\frac{r}{h}\right)^2\left(\frac{z}{h}\right)^2+3\left(1-\left(\frac{z}{h}\right)^2\right)^2}
\label{A2}
\end{equation}
where the origin of z-axis is placed in the center of symmetry plane and $h$ is half of the film thickness. Using the generalized shear rate as the plastic strain, the heat equation is now given in Eq. \ref{A3}
\begin{equation}
\frac{\partial\bar{T}}{\partial\bar{t}} = \frac{9\beta\eta{\dot{h}}^2}{4\alpha\rho c_pT_0}\left[{\bar{r}}^2{\bar{z}}^2+3(1-{\bar{z}}^2)^2\right]+\bar{{\nabla}}^2\bar{T}
\label{A3}
\end{equation}
where
\begin{equation}
\bar{T} = \frac{T}{T_0},\,\, \bar{r} = \frac{r}{h},\,\, \bar{z}=\frac{z}{h},\,\, \bar{t} = \frac{t}{(\frac{h^2}{\alpha})}
\label{A3sup}
\end{equation}
and
\begin{equation}
\bar{\nabla}^2 = \frac{1}{\bar{r}}\frac{\partial}{\partial\bar{r}}\left(\bar{r}\frac{\partial}{\partial\bar{r}}\right)+\frac{\partial^2}{\partial\bar{z}}
\label{A3sup2}
\end{equation}

Substituting the values of $\beta=1$ \cite{Mason1994}, and experimental parameters of $\eta=10^6$ Pa.s, $\alpha=10^{-5}$, $c_p\sim$ 500 J/(kg.K), $\rho$=15 g/cm$\mathrm{^3}$, $T_0$=543 K and taking the sample compressed at the maximum loading rate of 15 kN/min, $\dot{h}=5\,\mathrm{\mu m/s}$ we note that the dimensionless prefactor of the first term in Eq. \ref{A3} has a value of
\[
\frac{9\beta\eta{\dot{h}}^2}{4\alpha\rho c_pT_0}=\frac{9\times 1 \times 10^6 \times (5 \times 10^{-6})^2}{4\times 10^{-5} \times 543 \times (15\times 10^3) \times 500}=1.4\times 10^{-9} \ll 1
\]

As a result, we can neglect the first term on the right of Eq. \ref{A3}, i.e. the heat generation from plastic work in our squeezing flow can be neglected. In fact, for the squeezing thin film process, the heat conductance is dominated along z-axis, then the steady-state solution of Eq. \ref{A3} can be easily obtained and the temperature difference in the liquid film can be found within 1 K.

\section{Shear enhanced crystallization in Zr-based BMG}

\begin{figure}[h]
\includegraphics[width=80mm, scale=1]{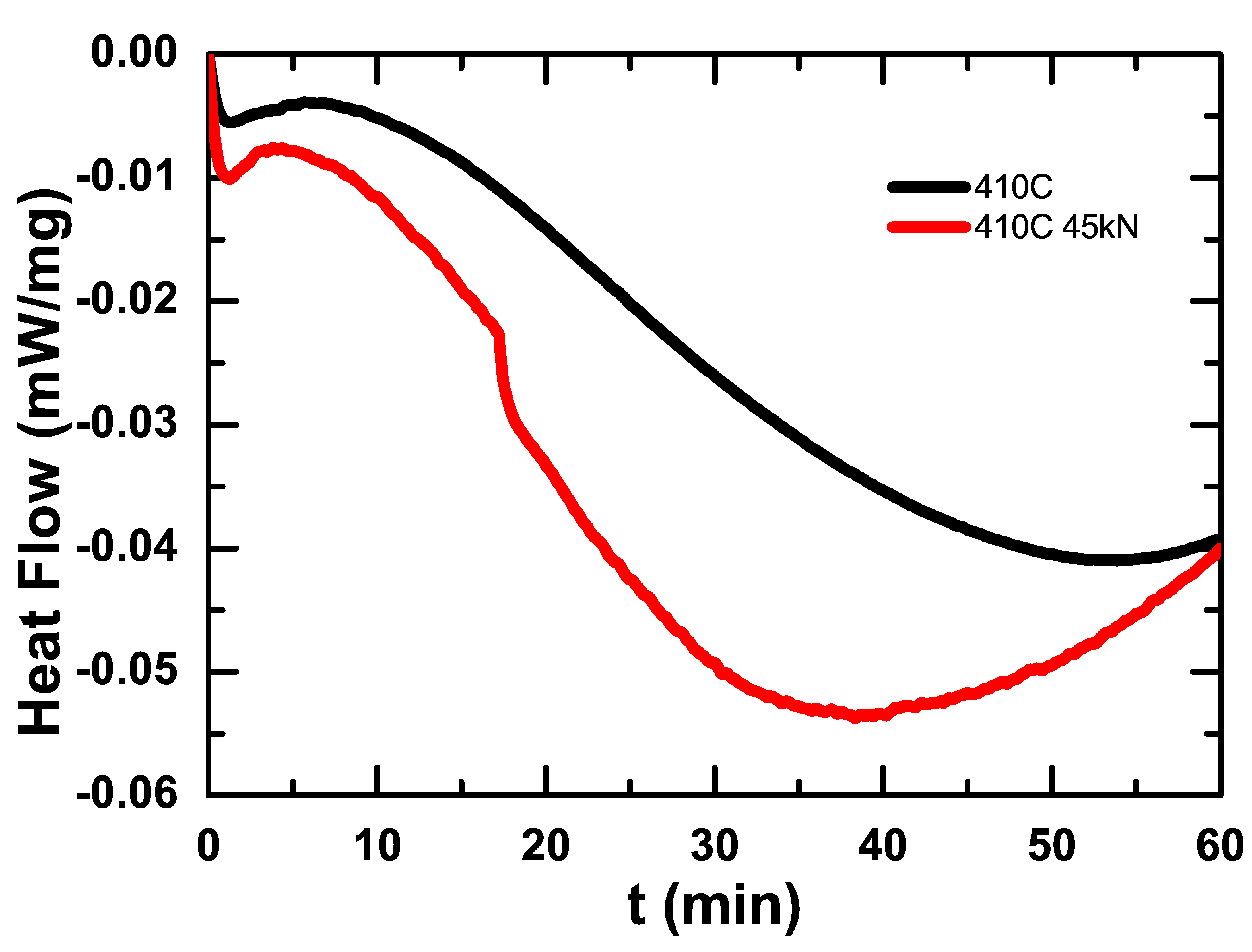}
\caption{Isothermal crystallization data at 410 $^{\circ}$C for Zr-based BMG. Data shown for a pristine sample and a disc pressed using a 45 kN/min loading rate.}
\label{Zr-BMG}
\end{figure}

A Zr-based BMG with composition $\mathrm{Zr_{44}Al_{10}Ti_{11}Cu_{10}Ni_{10}Be_{25}}$ with nominal $T_g$ and $T_x$ of 350 and 471 $^{\circ}$C was pressed at 410 $^{\circ}$C for 180 s using a loading rate of 45 kN/min. The sample was thereafter crystallized isothermally at 410 $^{\circ}$C to determine the crystallization time, relative to the undeformed material. Data are shown in Figure \ref{Zr-BMG}. The crystallization time is shortened from roughly 53 minutes to 36 minutes as a result of the processing.

\end{document}